\documentstyle[11pt,paspconf,psfig]{article}
\begin{document}

\title{Battery and Dynamo theory in the Kerr metric}

\author{Ramon Khanna}
\affil{Landessternwarte K\"onigstuhl,\\
D-69117 Heidelberg, Germany\\E-mail: Ramon.Khanna@lsw.uni-heidelberg.de}

\begin{abstract}
The generation and evolution of magnetic fields in the plasma 
accreting into a rotating black hole is studied in the 3+1 split of the Kerr 
metric. Attention is focused on effects of the gravitomagnetic potential. 
The gravitomagnetic force appears as battery term in the generalized 
Ohm's law. The gravitomagnetic battery is likely
to saturate at higher field strength than the classical Biermann battery.

The coupling of the gravitomagnetic potential with 
electric fields appears as gravitomagnetic current in
Maxwell's equations. In the magnetohydrodynamic induction equation, this
current re-appears as source term for the poloidal magnetic field, which can 
produce closed magnetic structures around an accreting black hole.
In principle, even self-excited axisymmetric dynamo action is possible,
which means that Cowling's anti-dynamo theorem does not hold in the Kerr 
metric.

Finally, simulations of the $\alpha\Omega$ dynamo in accretion flows 
into the hole are presented. 
I assume a simple expression of $\alpha$ in this relativistic context. 
The modes of the dynamo are oscillating for dynamo numbers which 
are typical for accretion disks. 
In a zero angular momentum flow into a Kerr black hole there is still 
shear, i.e. the angular velocity $\Omega$ of the plasma equals the angular 
velocity of space, $\omega$, and it has been speculated 
(Meier 1998) that even then a dynamo could operate. This is shown to be 
unlikely due to the rapid accretion. 
\end{abstract}

\section{Introduction}
The influence of the gravitomagnetic field of a rotating compact object 
on electromagnetic fields has been studied for some 25 years (Wald 1974, 
Ruffini \& Wilson 1975, Blandford \& Znajek 1977). 
The coupling of the gravitomagnetic potential with a magnetic field 
results in an electromotive force. 
Currents driven by this electromotive force may extract rotational 
energy from a black hole. This energy could power relativistic jets 
by Poynting flux (but see also, e.g., Punsly 1996).
Cast in the language of the 3+1 split of the Kerr metric, Maxwell's equations, 
together with the `ingoing wave boundary condition' for electromagnetic 
fields at the horizon, led to the {\it Membrane Paradigm} 
(Thorne et al. 1986).
Since a black hole does not carry its own magnetic field, not to mention 
kGauss fields required for the Blandford-Znajek process to be efficient, 
strong magnetic fields must either be accreted into the black hole 
from the outer accretion disk, or have to be generated and amplified 
in the plasma surrounding the black hole.

The generation of magnetic fields by a battery operating in the plasma 
close to a rotating black hole was studied by Khanna (1998b). 
It was shown that the gravitomagnetic force may play a crucial 
role in the battery. 
Khanna \& Camenzind (1996a) studied the possibility of 
an axisymmetric gravitomagnetic dynamo (or $\omega\Omega$ dynamo), 
in which the coupling between the 
gravitomagnetic potential and an electric field is a source for the 
poloidal magnetic field. This theoretical result invalidates 
Cowling's anti-dynamo theorem (see also N\`u\~nez 1997), 
but self-excited growing dynamo modes could not yet be numerically varified 
for simple kinematics (Khanna \& Camenzind 1996b). Egi et al. (1998) have 
given a criterion for growing modes of the $\omega\Omega$ dynamo, i.e. 
that the Poynting flux from close to the horizon 
(extracted rotational energy of the hole) be positive.

Recently, Meier (1998) speculated that, in a zero angular momentum flow 
into a Kerr black hole, alternatively to the $\omega\Omega$ dynamo, an 
$\alpha\omega$ dynamo might operate. My simulations of such 
a sceanrio show that, due to the relativistic accretion velocity, an 
$\alpha\omega$ dynamo is not very likely. 

Section \ref{MHD} gives an introduction to the derivation of MHD 
in the 3+1 split of the Kerr metric. The generalized  
Ohm's law for an electron-ion plasma is presented in Sec. \ref{gOhm}. 
The gravitomgnatic battery (along with the relativistic 
equivalent of Biermann's battery) are discussed in Sec. \ref{bat}.
In Sec. \ref{Indeq} the MHD induction equation in the 3+1 split 
of the Kerr metric is presented. Applications are the gravitomagnetic  
dynamo (Sec. \ref{gmdyn} and \ref{mfstruc}) and, in Sec. \ref{aOd}, 
the $\alpha\Omega$ dynamo, or the $\alpha\omega$ dynamo, respectively.
Throut the paper I set $G=1=c$.

\section{The MHD description of an electron-ion plasma}\label{MHD}
The formulation of MHD requires the relativistic definition of a plasma
as center-of-mass fluid of its components (Khanna 1998a).
The plasma is assumed to be a perfect fluid and is defined by the sum of the
ion and electron stress-energy tensors, which contain a collisional coupling 
term:
\begin{equation}
	(\rho_{\rm m}^{\, '} + p^{\, '})W^{\alpha}W^{\beta} + p^{\, '} 
        g^{\alpha\beta}
	\equiv T^{\alpha\beta} = \sum_{x=i,e}
	(\rho_{\rm mx}^{\rm x} + p_{\rm x}^{\rm x})
	W_{\rm x}^{\alpha}W_{\rm x}^{\beta} + p_{\rm x}^{\rm x} 
	g^{\alpha\beta} + T^{\alpha\beta}_{\rm x\, coll}\; .
\label{defplas}
\end{equation}
Subscripts $i,e$ refer to ion and electron quantities, respectively. 
Superscripts denote the rest-frame in which the quantity is defined, 
where $^{\, '}$ refers to the plasma rest-frame.
In the 3+1 split (into hypersurfaces of constant Boyer-Lindquist time $t$, 
filled with stationary zero angular momentum {\it fiducial observers})
\(T^{\alpha\beta} \) splits into the total density of mass-energy $\epsilon$ 
and momentum density $\vec{S}$
\begin{equation}
	\epsilon \equiv (\rho_{\rm m}^{\, '} + p^{\, '} v^2)\gamma^2
	\approx \rho_{\rm m}^{\, '}\gamma^2 
\qquad
	\vec{S} \equiv (\rho_{\rm m}^{\, '}+ p^{\, '})\gamma^2\vec{v}
	\approx \rho_{\rm m}^{\, '} \gamma^2\vec{v}
\label{epsSdef}
\end{equation}
and the stress-energy tensor of 3-space with metric 
$\buildrel\leftrightarrow\over{h}$
\begin{equation}
	\buildrel\leftrightarrow\over{T} \equiv (\rho_{\rm m}^{\, '}+
	p^{\, '})\gamma^2 \vec{v}\otimes\vec{v} 
	+ p^{\, '}\buildrel\leftrightarrow\over{h}
	\approx
	\rho_{\rm m}^{\, '}\gamma^2 \vec{v}\otimes\vec{v} 
	+ p^{\, '}\buildrel\leftrightarrow\over{h}\; .
\end{equation}
The approximate expressions hold for a `cold' plasma.
Charge density and current density are given by 
\begin{equation}
	\rho_{\rm c} \equiv \rho_{\rm ci} + \rho_{\rm ce} =
	Z e n_{\rm i}\gamma_{\rm i} - e n_{\rm e}\gamma_{\rm e}
\qquad
	\vec{j} \equiv \vec{j}_{\rm i} + \vec{j}_{\rm e} =
	Z e n_{\rm i}\gamma_{\rm i}\vec{v}_{\rm i}
	- e n_{\rm e}\gamma_{\rm e}\vec{v}_{\rm e}
    \; .
\end{equation}
All quantities resulting from the split are measured locally by FIDOs.
\subsection{The generalized Ohm's law in the 3+1 split of the Kerr metric}
\label{gOhm}
In the `cold' plasma limit, the local laws of momentum conservation for 
each species can be re-written as equations of motion, which can then be 
combined to yield the generalized Ohm's law for an electron-ion plasma
\begin{eqnarray}
	\frac{\vec{j}}{\sigma\gamma_{\rm e}}&\approx&
	\vec{E} +\frac{Z n_{\rm i}\gamma_{\rm i}}{n_{\rm e}\gamma_{\rm e}}
	\vec{v}\times\vec{B}
	-\frac{\vec{j}\times\vec{B}}{e n_{\rm e}\gamma_{\rm e}}
	+\frac{\vec{\nabla}(\alpha_{\rm g} p_{\rm e}^{\rm e})}
	{e n_{\rm e}\gamma_{\rm e}\alpha_{\rm g} }
	+\frac{4\pi\gamma_{\rm e}}{\omega^2_{\rm pe}}\rho_{\rm c}^{\; '}
	\vec{g}
	+\frac{\rho_{\rm c}{\; '}\gamma\vec{v}}{\sigma\gamma_{\rm e}}
	\nonumber\\
	&-&\frac{4\pi e(Z n_{\rm i}\gamma_{\rm e}^2  
	- n_{\rm e}\gamma_{\rm i}^2)}
	       {\omega_{\rm pe}^2 \gamma_{\rm e}\gamma_{\rm i}^2}
	\left(\frac{d(\gamma\vec{v})}{d\tau_{\rm p}}
       - \buildrel\leftrightarrow\over{H}\cdot(\gamma^2\vec{v})\right) \; ,
\label{allgOhm}
\end{eqnarray}
with the conductivity \( \sigma = {e^2 n_{\rm e}}/{m_{\rm e}\nu_{\rm c}}\equiv 
\omega^2_{\rm pe}/4\pi\nu_{\rm c}\) as measured in the plasma rest frame, 
the electron plasma frequency $\omega_{\rm pe}$, 
the factor of gravitational redshift $\alpha_{\rm g}$ 
(with \(\vec{g}=-\vec{\nabla}\ln\alpha_{\rm g}\)) and 
the gravitomagnetic tensor field \(\buildrel\leftrightarrow\over{H}\equiv 
\alpha_{\rm g}^{-1} \vec{\nabla}\vec{\beta}\, .\) 
\(\vec{\beta} = \beta^{\phi}\vec{e}_{\phi}\equiv -\omega\vec{e}_{\phi} \) 
is the gravitomagnetic potential, which drags space into differential 
rotation with angular velocity $\omega$. Note that, in the single fluid 
description, the 
gravitomagnetic force drives currents, only if the plasma is charged in its 
rest frame. $\tau_{\rm p}$ is the proper time in the plasma rest frame.
The derivation requires the assumption that the species 
are coupled sufficiently strong that their bulk accelerations 
\begin{equation}
	\frac{d(\gamma_{\rm x}\vec{v}_{\rm x})}{d\tau_{\rm x}}\equiv
	\left[\frac{\gamma_{\rm x}}{\alpha_{\rm g}}\frac{\partial}{\partial t}
	+\gamma_{\rm x}\left(\vec{v}_{\rm x}
	-\frac{\vec{\beta}}{\alpha_{\rm g}}\right)\cdot\vec{\nabla}\right]
	(\gamma_{\rm x}\vec{v}_{\rm x})
\label{bulkacc}
\end{equation}
are synchronized. The same is required for the gravitomagnetic accelerations, 
i.e. 
\(|\buildrel\leftrightarrow\over{H}\cdot(\gamma_{\rm i}^2\vec{v}_{\rm i}) -
\buildrel\leftrightarrow\over{H}\cdot(\gamma_{\rm e}^2\vec{v}_{\rm e})|
	 \ll
|\buildrel\leftrightarrow\over{H}\cdot(\gamma_{\rm i}^2\vec{v}_{\rm i})|
\). 
If the MHD-assumption of ``synchronized accelerations''
is not made, Ohm's law contains further current acceleration terms, 
inertial terms and gravitomagnetic terms (Khanna 1998a), which may be 
important for 
collisionless reconnection and particle acceleration along magnetic fields. 
This topic will be discussed elsewhere. 

In the limit of quasi-neutral plasma \((Z n_{\rm i}\approx  n_{\rm e})\) and
\(\gamma_{\rm e}\approx\gamma_{\rm i}\approx\gamma\), 
Eq.~(\ref{allgOhm}) reduces to
\begin{equation}
	\vec{j} \approx
	\sigma\gamma(\vec{E} +\vec{v}\times\vec{B}) 
	-\frac{\sigma}{e n_{\rm e}}(\vec{j}\times\vec{B})
	+\frac{\sigma}{e n_{\rm e}\alpha_{\rm g}}\vec{\nabla}(\alpha_{\rm g} 
	p_{\rm e}^{\rm e}) \; ,
\label{allgOhmqn}
\end{equation}
which contains all the terms, familiar from the non-relativistic generalized
Ohm's law, but no gravitomagnetic terms.
\subsection{The gravitomagnetic battery}\label{bat}
The generation of magnetic fields by a plasma battery was originally devised 
by Biermann (1950) for stars. He showed that, if the centrifugal force 
acting on a rotating plasma does not possess a potential, the charge 
separation owing to the electron partial pressure cannot be balanced 
by an electrostatic field, and thus currents must flow and a magnetic field 
is generated. 

In Khanna (1998b) I have re-formulated Biermann's theory in 3+1 split of 
the Kerr metric. The base of this battery theory is Ohm's law of 
eq.~(\ref{allgOhmqn}). Assuming that electrons and ions have
non-relativistic bulk velocities in the plasma rest frame, superscripts 
$i,e,^{\, '} $ can be dropped.
With \(p = p_{\rm i}+ p_{\rm e}= (n_{\rm i}+ n_{\rm e})kT\), the 
{\it impressed electric field} (IEF),
\(
	\vec{E}^{(i)} = {\vec{\nabla}(\alpha_{\rm g} p_{\rm e})}
	/{e n\gamma \alpha_{\rm g}}
\), 
can be re-expressed with the aid of the equation of motion for a 
`cold' quasi-neutral plasma to yield 
\begin{equation}
	\vec{E}^{(i)} = \frac{m_{\rm i}}{(Z+1)e }
	\left(\gamma\vec{g} + 
	\buildrel\leftrightarrow\over{H}\cdot(\gamma\vec{v})
	-{d(\gamma\vec{v}) \over d\tau}\right)
	+\frac{Z \left(\vec{j}\times\vec{B} 
	+ (\vec{v}\cdot\vec{j})\vec{E}\right) }
		{(Z+1) e n \gamma}\; .
\end{equation}
$\tau$ is the proper time in a FIDO frame; 
i.e. \(d / d\tau_{\rm p} = \gamma d / d\tau\).
The criterium for magnetic field generation is that 
$\vec{\nabla} \times{\alpha_{\rm g}\vec{E}^{(i)}}\ne 0\, .$ 
Here I restrict the discussion to 
the gravitomagnetic IEF $\vec{E}^{(i)}_{\rm gm}\; .$
The function part of \(\alpha_{\rm g}\vec{E}^{(i)}_{\rm gm}\) is
\begin{eqnarray}
	\lefteqn{
	\left(\vec{\beta}\cdot\vec{\nabla} + \vec{\nabla}\vec{\beta}\, 
	\cdot\right)
	(\gamma\vec{v}) =
	\left(\beta^i(\gamma v^j)_{|i} + \gamma\beta^{i|j} v_i\right)\vec{e}_j
	}\nonumber\\
	&&= -\gamma v^{\phi}{\tilde \omega}^2\vec{\nabla}\omega
	    -\omega\left((\gamma v^r)_{,\phi}\vec{e}_r
		+ (\gamma v^{\phi})_{,\phi}\vec{e}_{\phi}\right) \; ,
\label{Egm}
\end{eqnarray}
where ${\tilde \omega} = (h_{\phi\phi})^{1/2}$ and ${}_|$ denotes the 
covariant derivative in 3-space.
In axisymmetry \(\alpha_{\rm g}\vec{E}^{(i)}_{\rm gm}\) is clearly rotational,
unless some freak $\gamma$ should manage to make 
\(\gamma v^{\phi}{\tilde \omega}^2\) a function
of $\omega$ alone. Thus the gravitomagnetic force drives a 
poloidal current and generates a toroidal magnetic field. 
Only if $ v^{\phi}$ is non-axisymmetric, the gravitomagnetic IEF drives 
to\-ro\-idal currents. The total IEF \((\alpha_{\rm g}\vec{E}^{(i)}_{\rm gm}
+\alpha_{\rm g}\vec{E}^{(i)}_{\rm class})\) is 
likely to rotational in general. This will be quantified for specific 
velocity fields elsewhere.

In presence of a weak poloidal magnetic field the Biermann battery is 
limited due to 
modifications of the rotation law by the Lorentz force, rather than by 
ohmic dissipation. Then the contribution of the centrifugal 
force to the IEF becomes irrotational already at weak to\-ro\-idal fields 
(Mestel \& Roxburgh 1962). 
The gravitomagnetic battery term, on the other hand, 
is only linearly dependent on $\vec{v}$. The equilibrium field strength 
should therefore be higher than for the Biermann battery. 
\section{The MHD induction equation in the 3+1 split of the Kerr metric}
\label{Indeq}
In this section I review the axisymmetric dynamo equations in the 3+1 split 
of the Kerr metric (Khanna \& Camenzind 1996a). 
Ohm's law is assumed to be of the standard form for a 
quasi-neutral plasma; Hall-term and IEF are neglected. 
Combining Maxwell's equations (Thorne et al. 1986) 
with Ohm's law yields the MHD induction equation 
\begin{equation}
	\frac{\partial\vec{B}}{\partial t} = 
		\vec{\nabla} \times\left( (\alpha_{\rm g}\vec{v}\times\vec{B} ) 
	- \frac{\eta}{\gamma}
		\left(\vec{\nabla} \times(\alpha_{\rm g}\vec{B}) 
	+ (\vec{E}_{\rm p}\cdot\vec{\nabla}\omega)
		{\tilde \omega}\vec{e}_{\hat\phi}\right) 
	\right) + (\vec{B}_{\rm p}
	\cdot\vec{\nabla}\omega){\tilde \omega}\vec{e}_{\hat\phi}\; . 
\label{MHDindeq}
\end{equation}
The term standing with the magnetic diffusivity $\eta$ is the current density, 
which, via Amp\`ere's law, contains the coupling of the gravitomagnetic field 
with the electric field. In axisymmetry this is simply the shear of the 
poloidal electric field in the differential rotation of space, $\omega$. 
Another 
induction term is the shear of the poloidal magnetic field by $\omega$. This 
generates to\-ro\-idal magnetic field out of poloidal magnetic field 
even in a zero-angular-momentum flow.

\subsection{The gravitomagnetic dynamo}\label{gmdyn}
Introducing the flux $\Psi$ of the poloidal magnetic field and the poloidal 
current $T$
\begin{equation}
	\Psi = \frac{1}{2\pi}\int\vec{B}_{\rm p}\cdot d\vec{A} 
	= {\tilde \omega} A^{\hat\phi}
\qquad
	T = 2\int\alpha_{\rm g}\vec{j}_{\rm p}\cdot d\vec{A} 
	= \alpha_{\rm g}{\tilde \omega} 
	B^{\hat\phi} \; ,
\end{equation}
where $A^{\hat\phi}$ is the to\-ro\-idal component of the vector potential,
eq.~(\ref{MHDindeq}) splits into
\begin{eqnarray}
       \frac{\partial\Psi}{\partial t} &+& \alpha_{\rm g} 
	(\vec{v}_{\rm p}\cdot\vec{\nabla})\Psi
	-\frac{\eta{\tilde \omega}}{\gamma} 
	v^{\hat\phi}(\vec{\nabla}\omega\cdot\vec{\nabla}\Psi)
	-\frac{\eta{\tilde \omega}^2}{\gamma}\vec{\nabla}\cdot
	\left(\frac{\alpha_{\rm g}}{{\tilde \omega}^2}\vec{\nabla}\Psi\right)
	\nonumber\\ 
	&=&{} \frac{\eta{\tilde \omega}}{\gamma\alpha_{\rm g}}
	\left[ \left( T \vec{v}_{\rm p} -
	      \frac{\eta}{\gamma}\vec{\nabla} T\right)\times\vec{e}_{\hat\phi}
	\right]
	\cdot\vec{\nabla}\omega
\label{dtPsi}
\end{eqnarray}
\begin{eqnarray}
	\frac{\partial T}{\partial t} &+& \alpha_{\rm g} 
	(\vec{v}_{\rm p}\cdot\vec{\nabla})T
	 +\alpha_{\rm g}{\tilde \omega}^2 T 
	\left(\vec{\nabla}\cdot\frac{\vec{v}_{\rm p}}{{\tilde \omega}^2} 
	\right)
	 - \alpha_{\rm g} {\tilde \omega}^2 
	\vec{\nabla}\cdot\left(\frac{\eta}{\gamma{\tilde \omega}^2}
	\vec{\nabla} T\right)
	\nonumber\\ 
	 &=&\alpha_{\rm g} {\tilde \omega}(\vec{\nabla}\Psi\times
	\vec{e}_{\hat\phi}) 
	\cdot\vec{\nabla}\Omega
	\; . \label{dtT}
\end{eqnarray}
These equations are the relativistic equivalent of the classical axisymmetric 
dynamo equations. It is important to note, however, that {\it no mean-field 
approach} was made, but $\Psi$ has source terms anyway. They result from 
$\vec{E}_{\rm p}\cdot\vec{\nabla}\omega$. 
Obviously, Cowling's anti-dynamo theorem does not hold close to 
a rotating black hole. Growing modes of this gravitomagnetic dynamo were 
shown to exist for steep gradients of the plasma angular velocity $\Omega$ 
(N\'u\~nez 1997). According to Egi et al. (1998) growing modes require 
that the Poynting flux carrying rotational energy extracted from the hole be
positive.
For simple accretion scenarios, growing modes could not 
be found in kinematic numerical simulations (Khanna \& Camenzind 1996b). 
If, however, 
magnetic field is replenished by an outer boundary condition, the 
gravitomagnetic source terms generate closed loops around the black hole.
\subsection{The magnetic field 
structure in the accretion disk close to the hole}\label{mfstruc}
\begin{figure}[]
\psfig{width=\textwidth,figure=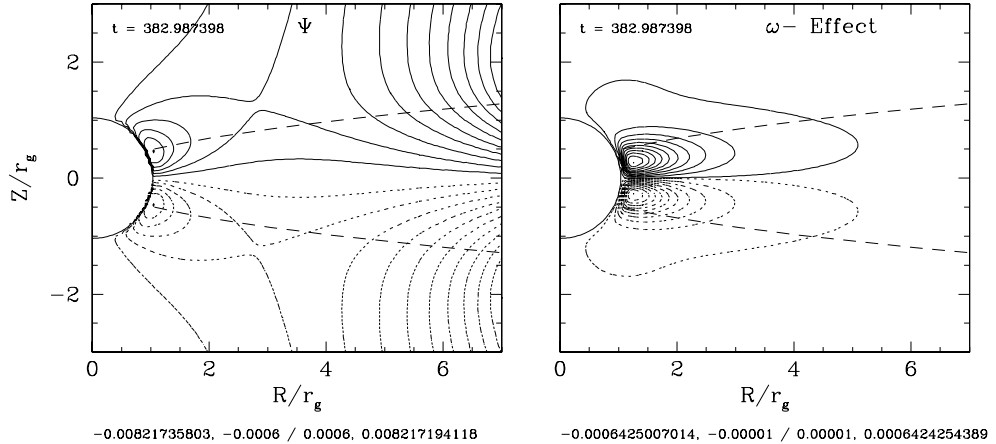,angle=-90}
\caption[ ]{Left: Contours of magnetic flux $\Psi$ showing a quadrupolar
magnetosphere of an accreting, rapidly rotating black hole ($a=0.998M$, 
$\alpha_{\rm visc} = 0.25$). 
Right: The gravitomagnetic current as source of $\Psi$.
The disk is marked by long-dashed lines. Solid contours correspond to 
positive values, short-dashed contours indicate negative values.
The range of contours is given below the boxes. Time is measured in diffusive 
timescales (see below). $r_{\rm g}=GM/c^2$.}
\label{gmdy}
\end{figure}

In an accretion disk, magnetic fields may be advected into the 
near-horizon area, where 
gravitomagnetic effects may become important. This can be simulated by 
advection/diffusion boundary conditions for $F=\Psi ,\ T$
\begin{equation}
	\frac{\partial F}{\partial n} + \frac{\gamma |v^{\hat r}|}{\eta}F = 
	\frac{\partial F_{\rm out}}{\partial n} 
	+ \frac{\gamma |v^{\hat r}|}{\eta}F_{\rm out}\; ,
\end{equation}
where $\partial /\partial n$ is the derivative along the outer boundary 
normal.
Figure~\ref{gmdy} shows the stationary final state of a time-dependent 
simulation (with turbulent magnetic diffusivity), in which 
$|B_{\rm p,out}| / |B_{\rm t,out}| = 1/50$. 
For such a dominantly to\-ro\-idal magnetic field 
the gravitomagnetic source terms are strong enough to change the 
topology of $\Psi$. This may influence the efficiency of the electromagnetic 
extraction of rotational energy from the hole.
\subsection{The {\boldmath$\alpha\Omega$} dynamo in the Kerr metric}
\label{aOd}
\begin{figure}[]
\psfig{width=\textwidth,figure=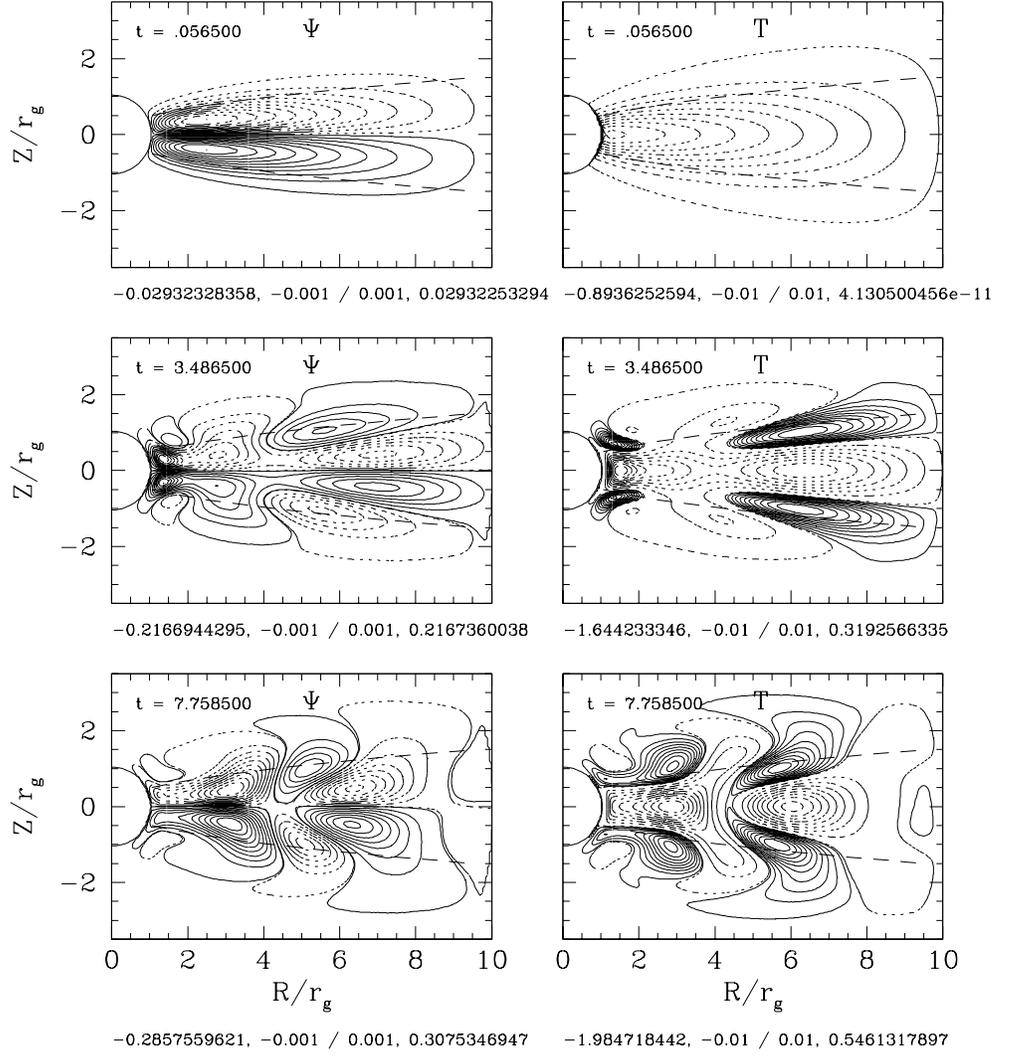,angle=-90}
\caption[ ]{A simulation of an $\alpha\Omega$ dynamo with symmetric initial 
current in a quasi-Keplerian accretion disk. 
Kerr parameter $a=0.998M$, $\alpha_{\rm visc}=0.065$. Solid contours have 
positive values, dashed contours have negative values. 
Kinks at the outer edge are artefacts of 
transforming data from the spherical grid into a carthesian plot. 
Simulation continued in Fig.~\ref{aO2}.
}
\label{aO1}
\end{figure}
\begin{figure}[]
\psfig{width=\textwidth,figure=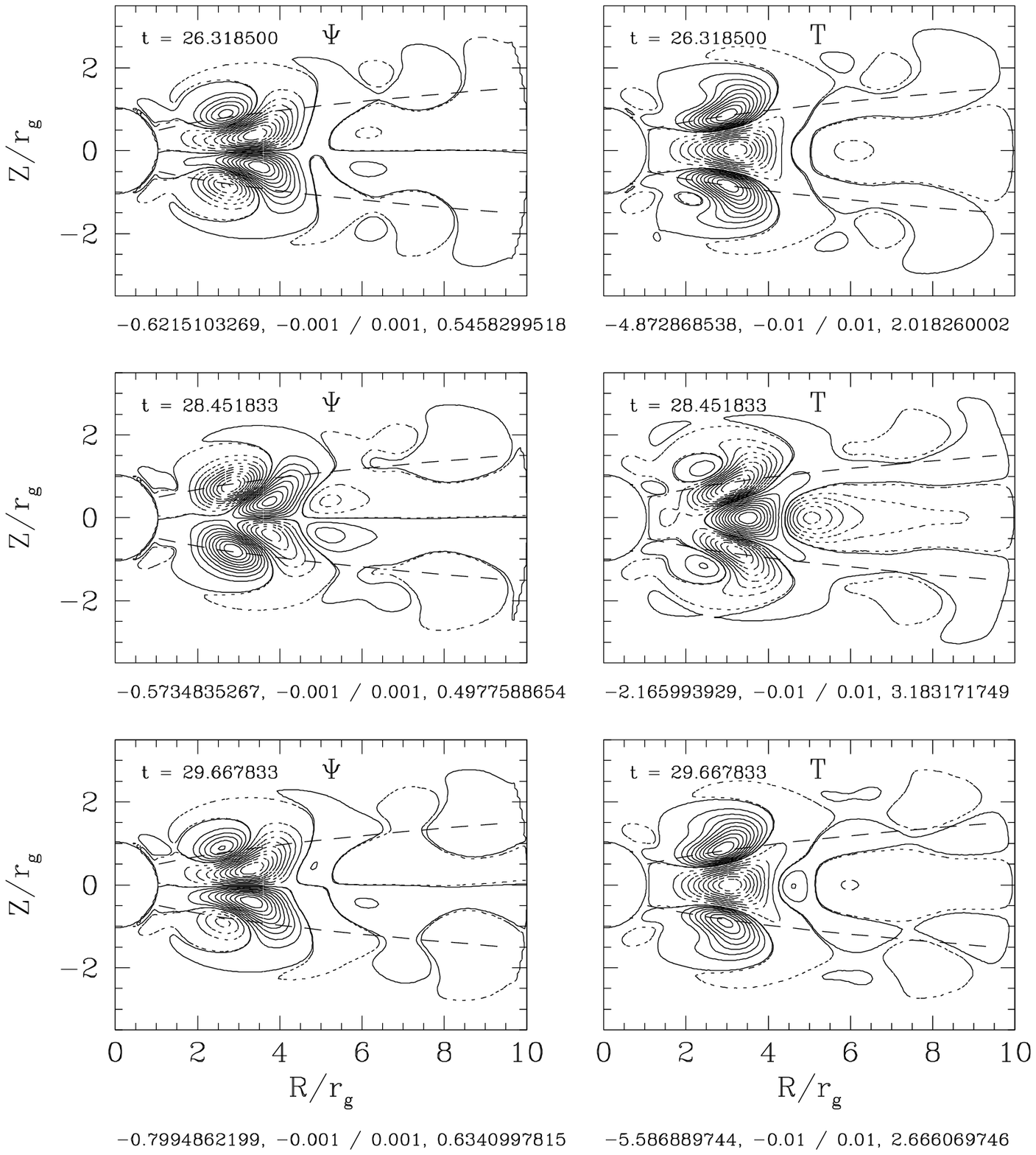,angle=-90}
\caption[ ]{The simulation of Fig.~\ref{aO1} has reached a slowly growing, 
oscillating eigenmode with a period of $\sim 3\, t_{\rm diff}$. Deviations from 
equatorial (anti-)symmetry are probably due to insufficient resolution in 
$\theta$-direction. Not shown here: The inclusion of non-linear 
$\alpha$-quenching leads to severe symmetry breaking and chaotic behavior.}
\label{aO2}
\end{figure}
In the innermost region of an accretion disk around a black hole there may 
also be a turbulent source term of $\alpha$-type. Without any knowledge of 
the physical source of the term (convection or magnetic shear instability) 
or its mathematical form in the relativistic context,
one can try to assess the physical regime (magnetic diffusivity, 
accretion velocity, rotation law) in which growing modes of an 
$\alpha\Omega$ dynamo exist. For a simple mean-field ansatz (Khanna \& 
Camenzind 1996a) the equations of the kinematic $\alpha\Omega$ dynamo 
are identical to Eqs.~(\ref{dtPsi}) and (\ref{dtT}) augmented by the 
$\alpha$-source term, $\alpha T$, for the flux and $\eta$ replaced by 
$\eta_{\rm turb}$. 
In analogy to the expression for $\alpha$ in classical disks, I assume
\begin{equation}
	\alpha = ( \alpha_{\rm g} R_o l_0^2 \Omega /H) \; f(z)/f(H/2) 
		= (3 \alpha_{\rm g} \alpha_{\rm visc} H \Omega)
		\; f(z)/f(H/2) \; ,
\label{adyn}
\end{equation}
where $H$ is the disk scale height, $R_o$ is the Rossby number and 
$\alpha_{\rm visc}$ is the viscosity parameter of standard accretion disk 
theory. The factor $\alpha_{\rm g}$ is added in order to suppress the 
source close to the horizon, where the accretion velocity approaches the 
speed of light (in properly derived mean-field equations there would 
probably be a ``Rossby number'' correlated to the accretion velocity instead). 
The vertical dependence of $\alpha$ is modelled with 
\(f(z) = \tanh(z) \exp[- (z/H)^2] \). 
The turbulent diffusivity is described as
\begin{equation}
	\eta_{\rm turb}=\alpha_{\rm visc} H^2 \Omega\, \tilde\omega 
	/ r\sin(\theta) \; ,
\end{equation}
with a vertical scaling \((\exp[- (z/H)^2] + 0.1)/1.1\).
The boundary condition at $r=10\,  r_{\rm g}$ is $\partial\Psi /\partial n 
= 0$ and $T=0$. 
Figure~\ref{aO1} shows the first part of a simulation with an initial 
current $T$, which is symmetric with respect to the equatorial plane, and 
$\Psi = 0$. The parameters are the Kerr parameter $a=0.998\, M$, 
$\alpha_{\rm visc}=0.065$, and the angular momentum of the accreting plasma 
is $99.999 \%$ of the Keplerian value at $r>r_{\rm ms}$ and constant within, 
which yields an accretion velocity of $\sim 0.003\, c$ at $r=3\, r_{\rm g}$, 
increasing to $c$ at the horizon. The dynamo is in a slowly growing 
quadrupolar mode, oscillating with a period of about three diffusive 
timescales \(t_{\rm diff} = r_{\rm g}^2 / \eta_0
\approx 2\ 10^5\; \sec\, M_9\, \left(\frac{\alpha_{\rm visc}}{0.1}\right)^{-1}\,
\left({H(r_{\rm h})\over 0.5 r_{\rm g}}\right)^{-2}\).

\begin{figure}[]
\psfig{width=\textwidth,figure=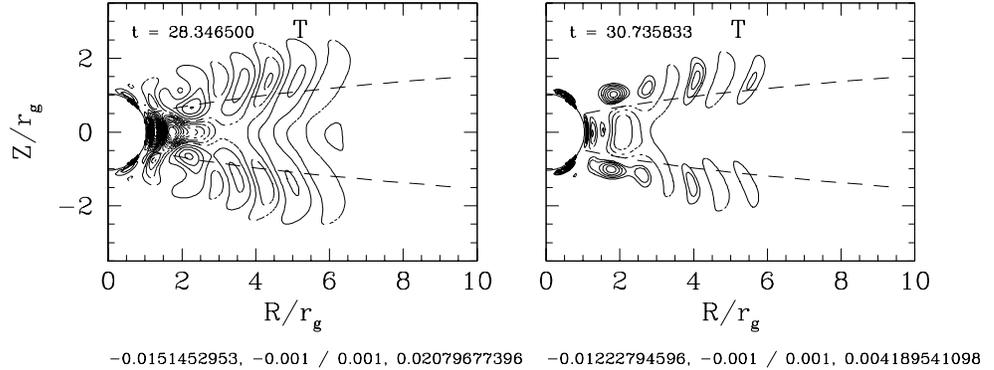,angle=-90}
\caption[ ]{Simulation with same setup as above, except that the accretion 
velocity at $r\ga 2\, r_{\rm g}$ is 10 times higher.
}
\label{aO3}
\end{figure}

The same setup, but with lower angular momentum in the accretion disk 
($99.9 \%$ of the Keplerian value), which corresponds to a radial velocity 
of $\sim 0.03 c$ at $r=3\, r_{\rm g}$, is in a decaying mode, 
which demonstrates that accretion impedes dynamo action (Fig.~\ref{aO3}).
	
\subsection{{\boldmath$\alpha\omega$} dynamo action in a zero-angular 
momentum flow?}\label{aod}
It was mentioned above that the shear of space does also induce a to\-ro\-idal
magnetic field 
(cf. eq.~[\ref{MHDindeq}]). In eq.~(\ref{dtT}) this shear term is 
obscure, but still there, hidden in 
\((\vec{\nabla}\Psi\times\vec{e}_{\hat\phi})\cdot\vec{\nabla}\Omega\propto
\vec{B}_{\rm p}\cdot\vec{\nabla} \Omega = \vec{B}_{\rm p}\cdot\vec{\nabla} 
(\alpha_{\rm g} v^{\phi}+ \omega)\). 
In a zero-angular-momentum flow $v^{\phi}=0$ (or, equivalently $\Omega = 
\omega$) 
and thus the current $T$ is solely generated by the shear of space. Moreover, 
$\vec{\nabla}\omega$ is significantly steeper than 
$\vec{\nabla}\Omega_{\rm K}$, which means that 
$\vec{B}_{\rm p}\cdot\vec{\nabla}\omega$ is a strong source term.

\begin{figure}[]
\psfig{width=\textwidth,figure=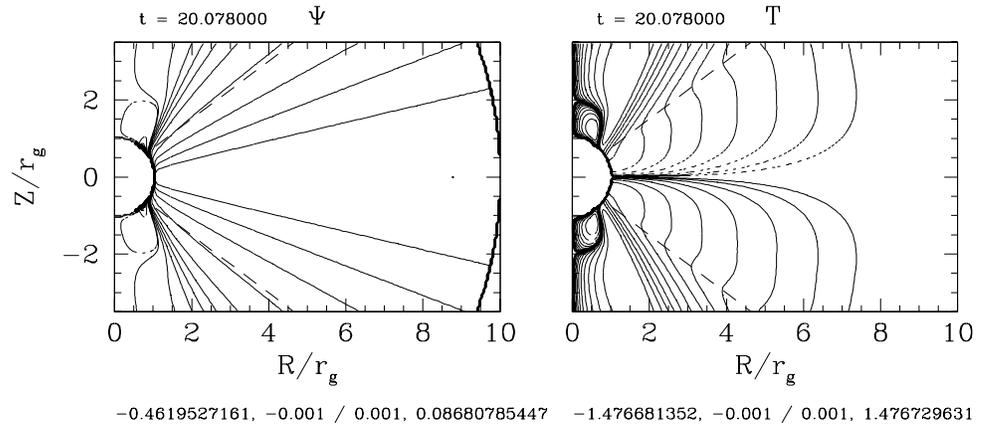,angle=-90}
\caption[ ]{Simulation of an $\alpha\omega$ dynamo. The dynamo generates 
transient structures in the corona close to the horizon. Within the 
disk there are no signs whatsoever of dynamo action. The field and current 
are completely determined by the relativistic advection. $T$ is shown in 
logarithmic contours.
}
\label{ao1}
\end{figure}

Meier (1998) speculated that, alternatively to the gravitomagnetic 
dynamo described above, there could be an $\alpha\omega$ dynamo in a 
zero-angular momentum accretion flow, with $\alpha$ being due to the magnetic 
shearing instability. Such a flow, however, accretes at relativistic velocities 
($\ga 0.1 \, c$ at $r=10\, r_{\rm g}$), which should 
suppress any dynamo action. This conclusion is supported 
by the simulation shown in Fig.~\ref{ao1}. The $\Psi$ and $T$ loops in 
the corona are transient and depend on the description of $\eta$ and 
$\alpha$ (here described as in Eq.~[\ref{adyn}], but 
not suppressed by $\alpha_{\rm g}$ in order 
to have an upper estimate of the source term). Parameters are 
$\alpha_{\rm visc}=0.003$ and $H(r_{\rm h})=0.8\, r_{\rm g}$. 
A wider parameter study is in progress.
\acknowledgments
This work was partly supported by the Deutsche For\-schungs\-ge\-mein\-schaft 
(SFB 328).


\begin{references}
\reference Biermann, L. 1950, Z. Naturforschg., 5a, 65
\reference Blandford, R.D., Znajek, R.L. 1977, MNRAS, 179, 433
\reference Egi, M., Tomimatsu, A., Takahashi, M. 1998, 
in: {\it The Central Regions of the Galaxy and Galaxies, 
IAU Symp. No. 184}, Y. Sofue (Ed.), Kluwer, p. 369
\reference Khanna, R. 1998a, MNRAS, 294, 673
\reference Khanna, R. 1998b, MNRAS, 295, L6
\reference Khanna, R., Camenzind, M. 1996a, AA, 307, 665
\reference Khanna, R., Camenzind, M. 1996b, AA, 313, 1028
\reference N\'u\~nez, M., 1997, Phys. Rev. Let., 79, 796
\reference Mestel, L., Roxburgh, I.W. 1962, ApJ, 136, 615
\reference Meier, D. 1998, astro-ph/9810352
\reference Punsly, B. 1996, ApJ, 467, 105
\reference Ruffini, R., Wilson, J.R. 1975, Phys. Rev. D, 12, 2959
\reference Thorne, K.S., Price, R.H., Macdonald, D.A., Suen, W.-M., 
Zhang, X.-H. 1986, in: {\it Black Holes: The Membrane Paradigm}, 
Thorne, K.S., Price, R.H., Macdonald, D.A., (Eds.), Yale Univ. Press, 
New Haven, pp. 67--120
\reference Wald, R.M., 1974, Phys. Rev. D, 10, 1680

\end{references}
\end{document}